\documentstyle[preprint,prc,aps,tighten]{revtex}
\begin{document}
\draft
\title{                  
 Next-to-Leading Order Description of Nucleon Structure Function In 
 Valon Model
}
\author{Firooz Arash$^{(a,b)}$\footnote{e-mail: arash@vax.ipm.ac.ir} and
Ali Naghi Khorramian$^{(c)}$}
\address{
$^{(a)}$
Center for Theoretical Physics and Mathematics, Atomic Energy
Organization of Iran, P.O.Box 11365-8486, Tehran, Iran \\
$^{(b)}$
Physics Department, Shahid Beheshti University, Tehran, Iran \\
$^{(c)}$
Physics Department, Amir Kabir University of Technology,
Hafez Avenue, Tehran, Iran \\
}
\date{\today}
\maketitle
\begin{abstract}
        We have improved and examined
the applicability of the {\it valon} model where the structure of any hadron is
determined by the structure of its constituent quarks. Nucleon structure 
functions are calculated within this model in the Next-to-Leading order. 
The results 
compare well with the experimental data. The model handles the bound state problem and the 
calculations show a flat or almost flat behavior for $F_{2}$ which sets in at some region of 
$x\leq 10^{-5}$at fixed $Q^{2}$. 
The emergence of this behavior is a consequence of the model and was not put 
in {\it a priori} as a theoretical guess. It seems that such a flatness 
can be inferred from HERA data, although, not completely confirmed yet. 
A set of parton distributions are given and their evolutions are tested. 
Some qualitative implications of 
the model for the spin structure of the proton is discussed. 
{\bf PACS Numbers  13.60 Hb, 12.39.-x, 13.88 +e, 12.20.Fv}
\end{abstract}
\maketitle
\newpage

\section    {INTRODUCTION} 
        Our understanding of hadrons is based on QCD for the interpretation
of Deep Inelastic Scattering (DIS) data together with the spectroscopic 
description 
of hadrons in terms of massive constituent quarks. At low energies static 
properties of the hadrons can be deduced from the latter. The most striking 
feature of 
the hadron structure intimately related to their nonvalence quark composition. 
This substructure can be generated in QCD,
however, perturbative approach to QCD does not provide absolute values of the 
observables and requires the input of non-perturbative matrix elements. 
Experimental data accumulated during the past ten years has shown that 
Gottfried Sum 
Rule is violated, suggesting strong violation of $SU(2)$ symmetry
breaking in the nucleon sea\cite{1}. It is possible to resolve 
the violation of Gottfried some rule by allowing a non-perturbative 
component to the nucleon sea\cite{2} or in a chiral quark model\cite{3}, in the region between 
the chiral symmetry-breaking scale $\Lambda_{\chi} \approx 1 GeV$ and the confinement 
scale $\Lambda_{QCD} \approx 0.1-0.3$ , hadron can be treated as weakly bound state of 
effective constituent quarks.The large violation of Ellis-Jaffe sum
rule implies that only a small fraction of proton helicity is carried by the
quarks, leading to the so-called "SPIN CRISIS". 
The process of quark evolution produces a large asymmetry for gluon, 
which at a scale of $10 (\frac{GeV}{c})^2$ can be quite large:$\Delta g\left[Q^{2}=10
(\frac{GeV}{c})^{2} \right] \approx 4$, and counter intuitive. In
fact, at this scale $ \Delta {g}$ can be anything; so, something has to
compensate for the spin component carried by the radiated gluon in the course
of evolution. It turns out that the orbital angular momentum of the produced
$q-\bar{q}$ and $q-g$ pairs is the compensating agent which finds a natural
place in the constituent picture of the nucleon; provided that a constituent
quark is viewed as an extended quasiparticle object composed of a valence quark
swirling with a cloud of gluons and sea $q-\bar{q}$ paris. All these fairly 
successful theoretical attempts 
suggest the presence of clusters in the nucleon. 
It seems that there is certain relationships between the constituent quark
model of hadron and its partonic structure.
Therefore, it is 
interesting to describe the nucleon structure in terms of these 
quasiparticle constituent quarks. The picture that emerges is as follows: 
At high enough $Q^{2}$ values it is the
structure of the constituent quark that is being probed and at 
sufficiently low
$Q^{2}$, no longer the constituent structure can be resolved.
        Of course, the above presented picture is not new. In 1974 Altarelli,
{\it{et al}} pioneered such a model, and more recently, pion structure
function
was constructed by Altarelli and co-workers in a constituent model\cite{4}. 
In
the early 1980's R.C. Hwa proposed a similar model, the so-called {\it{valon
model}} \cite{5} which was more elaborate and very successful in analyzing a
range of observed phenomena \cite {6} \cite {7}. 
The rise of $F_{2}$ at small-$x$ and larger scale of $Q^{2}\geq5 GeV^{2}$ 
posses a
Lipatov-type behavior, whereas, at smaller scale of $Q^{2}\approx 1-3 GeV^{2}$
it is predicted to develop Regge-type $x$-dependence;that is, parton
distributions and $F_{2}$ to be flat or almost flat \cite{8}. This latter
prediction although has not been confirmed yet at HERA, but there are some
evidences of flattening for $F_{2}$ at smallest $x$-bins in the HERA
data\cite {9} as well as EMC and NMC\cite{10} data.
        In this paper we will utilize the essence of the valon model of Ref.[5]
in an attempt to investigate the above mentioned qualitative regularities 
in the constituent quark model of hadronic structure in the HERA region which is now
extended to very low-$x$ values by H1 [9], \cite{11}, and ZEUS \cite{12} collaborations.
The organization of the paper is as follows; in section II we will give a brief
description of the valon model which this work is based on, then we will
calculate the nucleon structure in terms of the structure of the valons. In
section III, an explicit parameterization of the parton distributions and the
numerical calculations will be outlined; In section IV we discuss some 
qualitative implications of the model on the spin structure of the nucleon.
Finally, in section V, we will discuss the results and elaborate on the 
concluding results.

\section {THE VALON MODEL}
        The valon model is a phenomenological model which is proven
to be very useful in its applied to many areas of the hadron
physics. The main features of the model is as follows:
(detailed work can be found in Ref. [5] and references therein.)
        A valon is defined to be a dressed valence quark in QCD with a 
cloud of
gluons and sea quarks and antiquarks. Its structure can be resolved at high
enough $Q^{2}$ probes. The process of dressing is an interesting subject of its
own. In fact, progress is made to derive the dressing process in the context 
of
QCD \cite{13}. A valon is an effective quark behaving as a quasiparticle. In
the scattering process the virtual emission and absorption of gluon in a valon
becomes bremsstrahlung and pair creation, which can be calculated in QCD. 
At sufficiently low $Q^{2}$ the internal structure of a valon
cannot be resolved and hence, it behaves as a structureless valence quark. At
such a low value of $Q^{2}$, the nucleon is considered as bound state of three
valons, UUD for proton. The binding agent is assumed to be very soft gluons or
pions. The constituent picture of hadron
also can be based on the Nambu-Jona-Lasinio (NJL) model\cite{14} including the
six-fermion $U(1)$ breaking term. In this model, the transition to the partonic
picture is described by the introduction of chiral symmetry breaking
scale $\Lambda_{\chi}$ . Nevertheless, for our purpose, combination of all these
effects is summarized in finding the parton distributions in the constituent
quark and arriving at the nucleon structure functions. One subtle point is that
the valons, or constituent quarks, are not free, gluons are also needed. Thus,
in addition to valon degree of freedom, gluon degrees are also should be
considered. But, it is not known that in an infinite-momentum frame, the valons
carry all of the hadron momentum. This concern ultimately ought to be settled
by a reliable theory of confinement. As a working hypothesis we shall assume
that the valons exhaust the hadron momentum.
        Let us denote the distribution of a valon in a hadron by
$G_{\frac{v}{h}}(y)$ for each valon $v$. It satisfies the normalization 
condition
\begin{equation}
\int_{0}^{1}G_{\frac{v}{h}}(y) dy =1.
\end{equation}
and the momentum sum rule:
\begin{equation}
\sum_{v}\int_{0}^{1}G_{\frac{v}{h}}(y)y dy =1 
\end{equation}
where the sum runs over all valons in the hadron $h$.
Nucleon structure function $F^{N}(x,Q^{2})$ is related to the valon structure
function $f^{v}(\frac{x}{y},Q^{2})$ by the convolution theorem as follows;
\begin{equation}
F^{N}(x,Q^{2})=\sum_{v}\int_{x}^{1}dyG_{\frac{v}{N}}(y)f^{v}(\frac{x}{y},Q^{2})
\end{equation}
        We note that valons are the universal property of the hadron and
therefore its distribution is independent of the nature of the probe and
$Q^{2}$ value. As the $Q^{2}$ evolution matrix of $F_{2}$ does not depend on the
target, the $Q^{2}$ evolution of partonic density is the same for partons in a
proton or in a valon. It follows then, that if the convolution is valid at one
$Q^{2}$, it will remain valid at all $Q^{2}$. Notice that the moments of the
parton densities

\begin{equation}
{\cal{M}}(n,Q^{2})=\int_{0}^{1} dx x^{n-1}P(x,Q^{2})
\end{equation}
are simply given by the sum of products of the moments
\begin{equation}
M(n,Q^{2})=\sum_{v}{\cal{M}}_{\frac{v}{h}}(n){\cal{M}}(n,Q^{2})
\end{equation}
At sufficiently high $Q^{2}$, 
$f^{v}(\frac{x}{y},Q^{2})$ can be calculated accurately in the leading-order 
(LO) and Next-to-leading order(NLO) results in QCD and its moments are 
expressed in terms of the evolution
parameter defined by:
\begin{equation}
s=\it{ln}\frac{\it{ln}\frac{Q^{2}}{\Lambda^{2}}}{\it{ln}\frac{Q_{0}^{2}}
{\Lambda^{2}}}
\end{equation}
where $\Lambda$ and $Q_{0}$ are the scale parameters to be determined from the
experiments. From the theoretical point of view, both $\Lambda$ and $Q_{0}$
should depend on the order of the moments, however, in our approximation we
will take them independent of $n$. Since $f^{v}(z,Q^{2})$ is free of
bound state complications which are summarized in $G_{\frac{v}{h}}(y)$; we can
describe a U-type valon structure function, say $F_{2}^{U}$ as:
\begin{equation}
F^{U}_{2}=\frac{4}{9}(G_{\frac{u}{U}} +G_{\frac{\bar{u}}{U}}) +
 \frac{1}{9}(G_{\frac{d}{U}}+G_{\frac{\bar{d}}{U}}+G_{\frac{s}{U}}+G_{\frac
{\bar{s}}{U}})+...
\end{equation}
where $G_{\frac{q}{U}}$ are the probability functions for quarks and antiquarks
to have momentum fraction $z$ in a $U$-type valon at $Q^{2}$. Similar
expressions can be written for the $D$-type valon. Structure of a valon, then,
can be written in terms of flavor singlet(S) and flavor nonsinglet(NS)
components as:
\begin{equation}
F_{2}^{U}(z,Q^{2})=\frac{2}{9}z\left[G^{S}(z,Q^{2}) + G^{NS}(z,Q^{2})\right]
\end{equation}
\begin{equation}
F_{2}^{D}(z,Q^{2})=\frac{1}{9}z\left[2G^{S}(z,Q^{2}) -
G^{NS}(z,Q^{2})\right]
\end{equation}
For the electron or muon scattering $G^{S}$ and $G^{NS}$ in eqs.($8-9$) are
defined as:
\begin{equation}
G^{S}= \sum_{i=1}^{f}(G_{\frac{q_{i}}{v}}+G_{\frac{\bar{q}_{i}}{v}}) \hspace{1.5cm}
G^{NS}= \sum_{i=1}^{f}(G_{\frac{q_{i}}{v}}-G_{\frac{\bar{q}_{i}}{v}})
\end{equation}
For the neutrino and anti-neutrino scattering, similar relations can be
written, but we will not present them here since we are mainly concerned with
HERA data. In the moment representation we will have;
\begin{equation}
{\cal{M}}_{2}(n,Q^{2})=\int_{0}^{1}dx x^{n-2}F_{2}(x,Q^{2})
\end{equation}
\begin{equation}
{\cal{M}}_{\gamma}(n,Q^{2})=\int_{0}^{1}dx x^{n-1}G_{\gamma}(x,Q^{2})
\end{equation}
where, $\gamma =\frac{v}{N}$,$S$,$NS$. From these equations, for a nucleon
eq.($5$) follows:
\begin{equation}
M^{N}(x,Q^{2})=\sum_{v}M_{\frac{v}{N}}(n)M^{v}(n,Q^{2})
\end{equation}
Solution of the renormalization group equation of
QCD provides the moments of singlet and nonsinglet valon structure functions 
in the LO and NLO and they can be
expressed in terms of the evolution parameter of eq.($6$). These moments 
are given in \cite{15}, and \cite{16}.
To evaluate nucleon structure function we need the distribution of valons 
in a nucleon. To proceed, we take a simple form for the exclusive valon 
distribution
\begin{equation}
G_{UUD}(y_{1},y_{2},y_{3}) = N y_{1}^{\alpha} y_{2}^{\alpha} y_{3}^{\beta} 
\delta(y_{1} +y_{2} +y_{3} -1)
\end{equation}
The inclusive distribution can be obtained by double integration over 
unspecified variables. for example:
\begin{equation}
G_{U/p}(y)= \int dy_{2}dy_{3} G_{UUD}(y_{1},y_{2},y_{3})=B(\alpha +1,\alpha +\beta +2)^{-1}
y^{\alpha}(1-y)^{\alpha +\beta +2}
\end{equation}
The normalization factor is fixed by the requirement that
\begin{equation}
\int G_{U/p}(y)dy =\int G_{D/p}(y)dy =1
\end{equation}
The moments of these inclusive valon distributions are calculated according to 
equation (12) and they are given by
\begin{equation}
U(n)=\frac{B(a+n,a+b+2)}{B(a+1,a+b+2)} \hspace{1.5cm}
D(n)=\frac{B(b+n,2a+2)}{B(b+1,2a+2)}
\end{equation}
 where $B(i,j)$ is the Euler beta function with $a=0.65$ and $b=0.35$. After
 performing inverse Mellin transformation we get for the valon distributions:
 \begin{equation}
 G_{U/p}=7.98 y^{0.65} (1-y)^{2} \hspace{2cm} G_{D/p}=6.01 y^{0.35} (1-y)^{2.3} 
 \end{equation}
\section { PARTON DISTRIBUTIONS}  
        To determine the parton distributions, we denote their moments by
$M_{x}$(n,s), where the subscript $x$ stands for valence quarks $u_{v}$ and
$d_{v}$, as well as for the sea quarks, antiquarks and the gluon. The
moments are functions of the evolution parameter, $s$. They are given as,

\begin{equation}
M_{u}(n,s)_{v}=2U(n){\cal{M}}^{NS}(n,s) \hspace{1.5cm}
M_{d}(n,s)_{v}=D(n){\cal{M}}^{NS}(n,s)
\end{equation}
\begin{equation}
M_{sea}(n,s)=\frac{1}{2f}\left[2U(n)+D(n)\right]\left[{\cal{M}}^{S}(n,s)-{\cal{M}}^{NS}(n,s)\right]
\end{equation}
\begin{equation}
M_g=\left[2U(n)+D(n)\right]M_{gQ}(n,s)
\end{equation}
Where ${\cal{M}}^{S}(n,s)$ and ${\cal{M}}^{NS}(n,s)$ are the moments of the singlet and
nonsinglet valon structure functions and $M_{gQ}(n,s)$ is the quark-to-gluon 
evolution function. $U(n)$ and $D(n)$ are the $U$ and $D$ type valon
moments, respectively, and are given in the previous section.
Calculation of $M_{x}$ is simple; in what follows, instead, the results
are presented in parametric form in the NLO.\\
        In determining the patron distributions, we have used a procedure
which is consistent with our physical picture. To assure that we are at a high
enough $Q^{2}$ value, we first used a set of data at $Q^{2}=12GeV^{2}$ 
from 1992 run of H1 collaboration which covers $x$ interval of
$x=[0.000383,0.0133]$. That is, for a single value of $s$,
or $Q^{2}$, we fit the moments by a sum of beta functions that are the 
moments of the form:
\begin{equation}
xq_{v}(x)=a(1-x)^{b}x^{c} \hspace{1.5cm}
xq_{sea(gluon)}(x)=\sum_{i}^{3}a_{i}(1-x)^{b_{i}}
\end{equation}
The fit is effective as seen in Figure 1. The parameters $a$,$b$,$c$, $a_{i}$, 
$b_{i}$, are further considered to be functions of $s$, the evolution 
parameter. For the valence sector :
\begin{equation}
a=a_{0} +a_{1} s +a_{2} s^{2}
\end{equation}
similarly for b, and c.
For the non-valence sector we have
\begin{equation}
a_{i}=\alpha_{i} + \beta_{i} exp(s/\gamma_{i})
\end{equation}
with similar form for $b_{i}$. The values for these parameters are given 
in the appendix. Since the formalism has to include both valence and the 
sea quarks, it is further
required that the valence distributions to satisfy the normalization
conditions:
\begin{equation}
\int_{0}^{1}q_{u_{v}}(x,Q^{2})dx=2 \hspace{1.5cm}
\int_{0}^{1}q_{d_{v}}(x,Q^{2})dx=1
\end{equation}
at all $Q^{2}$ values, reflecting the number of valence quark of each type.
For the sea quarks distributions we will assume $SU(2)$ flavor symmetry
breaking inferred from the violation of the Gottfried sum rule \cite{1}.
Implementation of the $\bar{u}<\bar{d}$ in the nucleon sea is those
obtained in Ref.[7] where we extracted the ratio:
\begin{equation}
\frac{\bar{u}}{\bar{d}}=(1-x)^{3.6}
\end{equation}
using low $p_{T}$  physics in the valon-recombination model. We take
$xs(x)=\frac{x(u_{sea}(x)+d_{sea}(x))}{4}$ and 
$xc(x)=\frac{1}{10}x(u_{sea}(x)+d_{sea}(x))$. These choices are made based 
on the mass ratios of the involved quarks. All distributions are referring 
to the proton. The parameterization of patron distributions are those given 
in equations (22-24).  From the fit to the data at 
$Q^{2}=12 GeV^{2}$ we determined the scale parameter $Q_{0}^{2}$ and $\Lambda$:
\begin{equation}
Q_{0}^{2}=0.28 GeV^{2} \hspace{2cm}
\Lambda=0.22 GeV.
\end{equation}
Figure 2 shows the shape of the distributions in equations (22) for
the valence and the sea quarks at typical values of $Q^{2}$ . 
Our sea quark distribution, although
has a sharp rise at small-$x$, but also damped very fast as $x$ increases. 
It appears that there are
some evidences in the H1 data at low-$x$ bins and low but fixed $Q^{2}$
that supports a flat shape for $F_{2}(x)$. Figure 3 presents this behavior. Such
a flattening of $F_{2}$ also is elucidated by Gluck, Reya and Vogt in Ref.
[8]. Our results favor a somewhat flat or almost
flat behavior which sets in at some $x_{0}\leq10^{-5}$ for low $Q^{2}$. This
is attractive in the sense that one may argue that the observed flatness
corresponds to the small-$x$ Regge behavior.
It is possible to generate the rise of $F_{2}$ as $x$ decreases either from
the DGLAP evolution in $\it{ln}Q^{2}$ or from BFKL evolution in
$\it{ln}(\frac{1}{x})$; although, due to large partonic densities at low $x$;
they both must be modified to account for the parton recombination effects.
Fortunately, at extreme limit of $x\rightarrow0$, behavior of parton
densities can be calculated analytically, though this limit will not be
reached within the kinematic range accessible to HERA.
It is evident from Figure 3 , that our calculation of the proton structure
function at low $Q^{2}$; $Q^{2}\geq0.4$; and small $x\approx10^{-5}-10^{-4}$
gives good agreement with the HERA data. we have included GRV(94) results of  
\cite{17} for comparison. In figure 4, $F_{2}$ is plotted as a function of $x$ 
for various $Q^{2}$ values.
In Figure 5, $F_{2}$ data is plotted
as a function of $Q^{2}$ for different $x$ bins. The data now extended over
four orders of magnitude both in $x$ and $Q^{2}$ and our constituent model NLO 
calculation
agrees well with experimental data. In our calculation for $Q^{2}<4GeV^{2}$ 
we have considered
only three flavors whereas for higher $Q^{2}$, four active flavors are taken
into consideration.
Since our main input in determining parton distribution was HERA data at
$Q^{2}=12 GeV^{2}$; and it is limited to low $x$ interval 
$[0.000383,0.0133]$;
it requires to check that if a satisfactory result for $F_{2}$ emerges for
the entire range of $x$. This is presented in Figure 6 for
$Q^{2}=20 GeV^{2}$ and $x=[0.000562,0.875]$ with the combined data
from H1 Ref.[11], BCDMS, SLAC and EMC taken from the compilation in
reference \cite{18}.
Functional form of gluon distribution is treated similar to the sea quarks
distribution as give in equation (22-24). The pertinent
parameters are obtained by imposition of the momentum Sum rule. Unfortunately,
there are only a few experimental data points for the gluon distribution,
to check against. These data points are the result of a direct measurement of
gluon density in the proton at low $x$ \cite{19}. Figure 7, shows the
accuracy of our gluon density and the data of Ref.[19], also the GRV
results are shown. Finally in figure 8 we plot $\frac{d_{v}}{u_{v}}$ for our
model and compare it with the world data.
Our result for the Gottfried sum rule with four flavors gives
\begin{equation}
S_{G}=\int_{0}^{1}\frac{dx}{x}\left[F_{2}^{p}-F_{2}^{n}\right] =0.27.
\end{equation}
Exactly the same result is obtained from the latest MRST parameterization 
\cite{20}
which is slightly larger than the experimental data of $S_{G}=0.235\pm 0.026.$ This discrepancy 
is originated from the form of $\frac{\bar{u}}{\bar{d}}$ used in eq.(26).
The point here is to demonstrate that a constituent quark model is able to 
describe the DIS data, even without a fine tuning.
\section {IMPLICATION ON THE NUCLEON SPIN} 
Spin structure of the nucleon merits a separate consideration of its own. 
Here we will mention a couple points relevant to the present work. \\ 
(i) In the naive quark model the polarized structure function $g_{2}$ is zero.
However, if we allow quarks to have an intrinsic $p_{\perp}$ inside the nucleon 
we can achieve a nonzero value for $g_{2}$:
\begin{equation}
g_{2}(x)=\frac{1}{2}\sum_{q}e^{2}_{q}(\frac{m_{q}}{xM}-1)\Delta q(x)
\end{equation}
with obvious notations. Neglecting the $p_{\perp}$ component leads to 
$m_{q}=xM$
and hence,$g_{2}=0$ is recovered. In parton model, even such an allocation of
$p_{\perp}$ and getting a nonzero value for $g_{2}$ is not free of ambiguity.
For, the parton model assumes the validity of impulse approximation and 
neglects the binding effects of the struck parton in large transverse 
momentum reactions. Measurement of polarization asymmetries may reveal that 
they depend on the binding energy. In the model described above, the 
binding effects are summarized in the constituent quark distributions in the 
nucleon and the structure of a constituent quark is free of binding 
problem. In this model $p_{\perp}$ distribution of quarks within the 
constituent quark can be obtained by realizing that there exists   
a size hierarchy: hadron size, constituent quark size, and the point-like 
partons as stated in\cite{21}. The hadronic structure is determined by the 
constituent quark wave function and its size is related to low-$Q^{2}$ 
form factor of the nucleon. Constituent quarks have a smaller size and 
described by their own form factors. Partons are the contents of the 
constituent quarks and are manifested only in high-$Q^{2}$ reactions. 
In high energy collisions one encounters two scales in $p_{\perp}$ 
distribution: the 
average transverse momentum of pions in multiparticle production processes 
is about $0.35$ GeV whereas, in massive lepton pair production
one needs to give a primordial $<p_{\perp}>\approx 0.8$ Gev to the partons 
in order to describe the data. These two scales are related to the 
hierarchy of sizes. Pion production is a soft process and its scale is 
related to the average transverse momentum of the constituent quarks, 
characterizing the hadronic size. Lepton pair production in $q-\bar{q}$ 
annihilation is a hard process and its scale is due to the transverse 
momenta of partons in the constituent quark. Using lepton pair production
data, the transverse 
momentum distribution of quarks in a constituent quark can be parameterized 
in a Gaussian form
\begin{equation}
p_{\perp q}^{c} (k^{2})=exp(-1.2 k^{2})
\end{equation}
which leads to the ratio of the sizes $\frac{<r^{2}>_{c}}{<r^{2}>_{hadron}}\approx\frac{1}{5}$.
Now we can calculate $<L_{z_{q\bar{q}}}>$. For a proton of radius $0.85 \it{fm}$
$$<L_{z_{q\bar{q}}}>=r_{c}<k_{\perp}>=0.321$$  
(ii) Another point related to the spin content of proton is also relevant 
here. H. Kleinert\cite{22} was first who suggested to consider the 
vacuum of massive constituent quark as a coherent superposition of the 
{\em Cooper pairs} of massless quarks in analogous to the theory of 
superconductivity. In the BCS theory, gauge symmetry associated with the 
particle number conservation is spontaneously broken and the Noether current 
$j^{\mu}_{5}$ is not conserved. To restore it,it is realized that there 
are both collective and single particle excitations. Recently Gaitan\cite{23} 
has shown that the bare current $j^{\mu}_{5}$ becomes dressed by a virtual 
cloud of Goldstone excitations ($q-\bar{q}$ in our case) and the conserved 
dressed current $j^{\mu}$ is the sum of two parts
\begin{equation}
j^{\mu}=j^{\mu}_{s}+j^{\mu}_{back}
\end{equation}
where $j^{\mu}_{back}$ describes the backflow current. This is very similar 
in our model, where for the dressed constituent quark, the generator of the 
gauge transformation induces a rotation of the $q\bar{q}$ pair  
correlations which can be identified as the orbital angular momentum. To 
this end, the pairing correlation will have the axial symmetry around an 
anisotropic direction, acting as the local $z$-axis and the particles forming 
the cloud of the constituent quark would rotate about this anisotropic 
direction. what have been said was only a qualitative description; to make 
it more quantitative, let us consider the spin of a constituent quark,say 
$U$. It can be written as
\begin{equation}
J^{U}_{z}=\frac{1}{2}=\frac{1}{2}\Delta\Sigma^{U}+L_{z_{q\bar{q}}}
\end{equation}
The DIS polarization data suggest that $\Delta\Sigma^{p}\approx\frac{1}{3}$.
Within the $SU(6)$ model\cite{24}
\begin{equation}
\Delta\Sigma^{p}=(\Delta U+\Delta D)\Delta\Sigma^{U}=\Delta\Sigma^{U}
\end{equation}
comparing with equation (32) we see that
$L_{z_{q\bar{q}}}\approx \frac{1}{3}$
i.e. about $70$ percent of the spin of a constituent quark is due to the 
orbital angular momentum of quark pairs in its surrounding cloud, screening the 
spin of the valence quark:
\begin{equation}
\frac{1}{2}\Delta\Sigma^{U}=S_{u_{val.}}+S_{sea}=\frac{1}{2} 
+S_{sea}=\frac{1}{6}
\end{equation}
resulting in $S_{sea}=-L_{z_{q\bar{q}}}=\frac{1}{3}$. We can take the 
transverse momentum distribution from eq.(30) and calculate $<L_{z_{q\bar{q}}}>$ 
directly. For the proton of mean radius of $0.86 fm$ we will have the mean 
radius of the constituent quark equal to $0.385 fm$ and hence,
$$<L_{z_{q\bar{q}}}>=r_{c}<k_{\perp}>=0.321$$
which agrees well with the results stated above. Notice that our calculation 
of $J_{U}$ and the conclusion reached did not include gluonic effects. In 
reality we should have included the gluon degree of freedom, however, an 
inclusion of those effects 
at the constituent quark level would reduce ($\Delta U+\Delta D$)by some 
$30$ percent from unity \cite{24}, nevertheless the point is clear.\\
\section {CONCLUSION} 
We have used the valon model to describe the deep inelastic scattering. The
model handles the bound state problem and sets the scale parameters. In
determining the parton distributions no arbitrary theoretical assumptions 
are made for low $Q^{2}$. Our parton distributions nicely accommodate the data 
in a wide range of $x=[10^{-6},1]$ and in a broad range of $Q^{2}$ spanning 
from a few $GeV^{2}$ up to $5000 GeV^{2}$. Our findings indicate that at 
$x\leq 10^{-5}$ at fixed $Q^{2}$ the structure function $F_{2}$ flattens, 
which may be interpreted as the manifestation of Regge dynamics. As 
$Q^{2}$, increases, however, the almost flat shape of $F_{2}$ gets washed out 
or pushed towards yet smaller region of $x$. Thus, we conclude that in a 
region of $x-Q^{2}$ plane the Regge dynamics is in play. This region sets in 
at some $x_{0}$ and not too high $Q^{2}$. Scale violation is obvious from the 
calculations and the data. The rise of $F_{2}$ at $Q^{2}<1$ $GeV^{2}$ in our 
model indicates that even at $Q^{2}$ as low as a few GeV the evolution has 
run the course. We further find that certain issues related to the spin
structure of the proton can find an interesting place in the framework of the 
model described.
\section {Appendix}
In this appendix we give the numerical values for our parton distributions
in the NLO both for three and four flavors. These relations are functions of 
evolution parameter $s$ defined in equation $(6)$. For details see the text. \\
$f=4$: \\
{\it{i}}: For valence sector using Eqs.(22-23) we have: \\
$q_{v}=u_{v}$ \\
$a=14.132-15.759s+6.795s^2-1.082s^3$, \\
$b=1.608-1.206s+0.37s^2-0.0523s^3$,  \\
$c=2.083+0.753s+0.214s^2-0.0106s^3$, \\
$q_{v}=d_{v}$ \\  
$a=14.509-5.309s+2.795s^2-0.562s^3$,  \\
$b=1.235-1.063s+0.467s^2-0.086s^3$,   \\
$c=2.215+0.639s+0.439s^2+0.02s^3$,  \\
{\it{ii}}: The sea distribution is parametrized as in Eqs. (22-24) with the following values.
Antiquark distributions are followed from  Eq.(26) and note thereafter. \\
$a_1=0.202+0.045\exp (s/0.791)$, \\
$b_1=3.09+4.187\exp (s/0.538)$, \\
$a_2=-0.064+0.152\exp (s/0.89)$ , \\
$b_2=2.574+11.156\exp (s/0.303)$, \\
$a_3=-0.196+0.014\exp (s/0.404)$, \\
$b_3=3.809+1.856\exp (s/0.669)$,  \\
{\it{iii}}: The gluon distribution  is given in Eqs.$(22-24)$ with the following 
numerical values. \\
$a_1=-2.81+0.46\exp (s/0.349)$, \\
$b_1=-4.565+8.319\exp (s/0.498)$, \\
$a_2=-0.948+0.297\exp (s/1.099)$, \\
$b_2=1.493+1.291\exp (s/0.809)$, \\
$a_3=-0.481+0.721\exp (s/0.948)$ , \\
$b_3=0.22+1.526\exp (s/1.306)$, \\\\
$f=3$: \\
{\it{i}}: For valence sector using Eqs.(22-23) we have: \\
$q_{v}=u_{v}$ \\
$a=10.891-12.525s+4.87s^2$, \\
$b=1.457-1.18s+0.384s^2$, \\
$c=1.979+0.254s+0.537s^2$, \\
$q_{v}=d_{v}$ \\ 
$a=5.977-7.55s+3.025s^2$, \\
$b=1.095-0.44s-0.067s^2$, \\
$c=2.532+0.282s+0.473s^2$, \\
{\it{ii}} The sea distribution is parametrized as in Eqs. (22-24) with the following values.
Antiquark distributions are followed from  Eq.(26) and note thereafter. \\
$a_1=0.31+0.064\exp (s/0.807)$, \\
$b_1=3.905+26.187\exp (s/0.658)$, \\
$a_2=-0.064+0.023\exp (s/0.305)$, \\
$b_2=2.644+14.156\exp (s/0.152)$, \\
$a_3=-0.063+0.023\exp (s/0.509)$, \\
$b_3=3.508+12.856\exp (s/0.452)$, \\
\begin{figure}
\label{fig1}
\caption{
 $F^{p}_{2}$ as a function of $x$ at $Q^{2} = 12 GeV^{2}$. Solid
line is our results in NLO, dashed line is that of GRV(NLO). Data are from 
Ref.[11,12]}
\end{figure}
\begin{figure}
\label{fig2}
\caption{
Model calculation of parton distributions at various $Q^{2}$.}
\end{figure}
\begin{figure}
\label{fig3}
\caption{
Variation of $F^{p}_{2}$ as a function of $x$ for $x < 10^{-4}$.}
\end{figure}
\begin{figure}
\label{fig4}
\caption{
$F^{p}_{2}$ as a function of $x$ for different $Q^{2}$ values. Solid
line is our results in NLO, dashed line is that of GRV(NLO). Data are from 
HERA.}
\end{figure}
\begin{figure}
\label{fig5}
\caption{
$F^{p}_{2}$ as a function of $Q^{2}$ for different $x$-bins. Solid
line is our results , dashed line is that of GRV(NLO). Data are from HERA}
\end{figure}
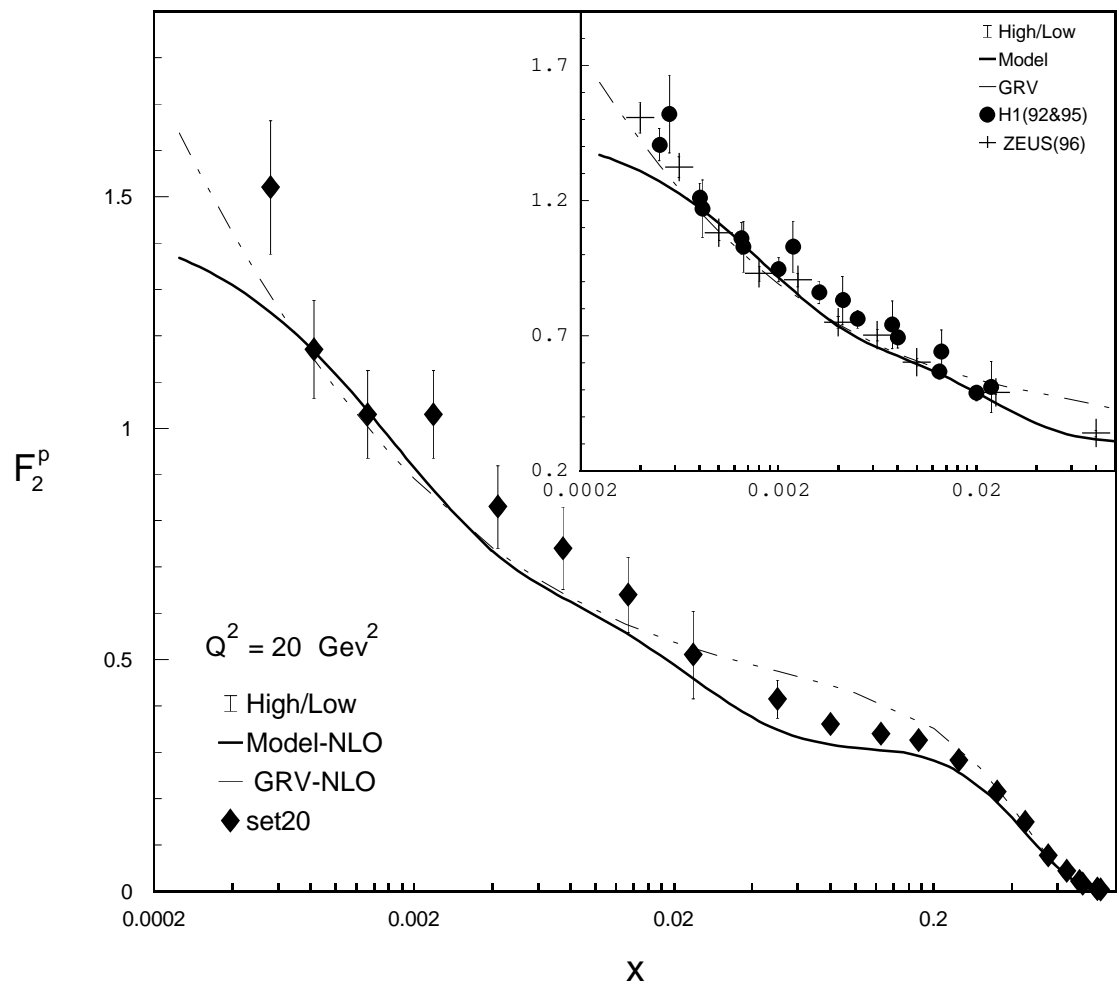
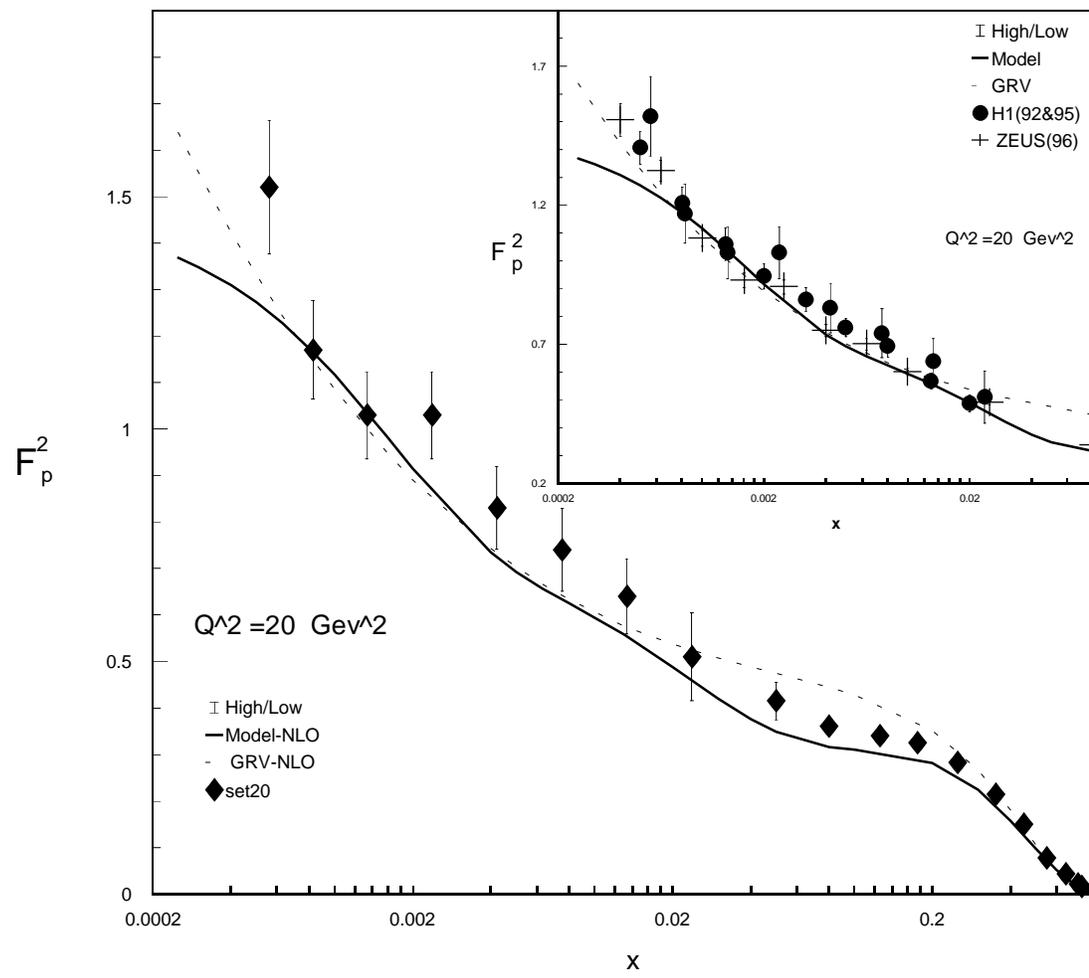
\begin{figure}
\label{fig6}
\caption{
$F^{p}_{2}$ at $Q^{2}=20 GeV^{2}$ for the entire range of $x$. Data are from
Refs.[11,12,18]}
\end{figure}
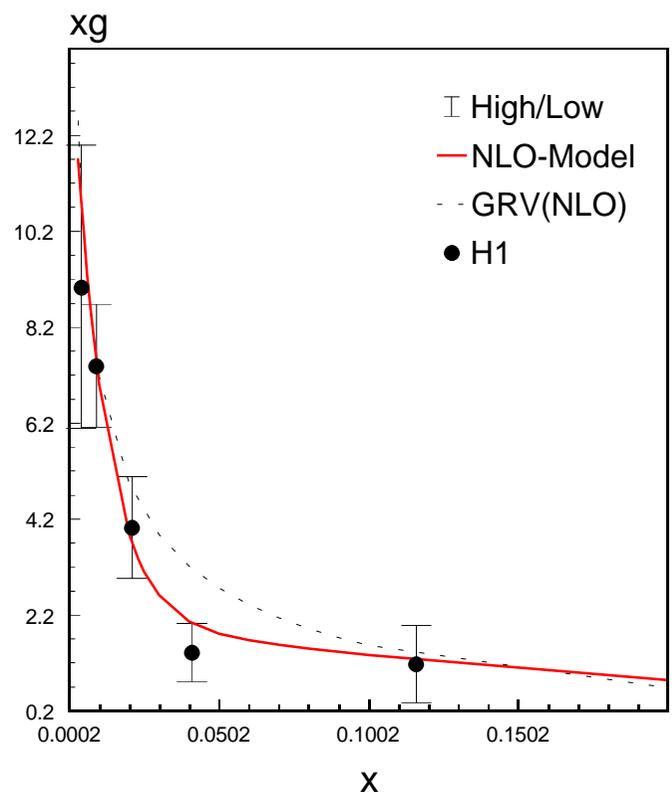
\begin{figure}
\label{fig7}
\caption{
$xg(x,Q^{2})$ at $<Q^{2}> = 20 GeV^{2}$ and $f=4$. Solid line is our results in NLO,
dashed line is that of GRV(NLO).  Data are from 
Ref.[19]}
\end{figure}
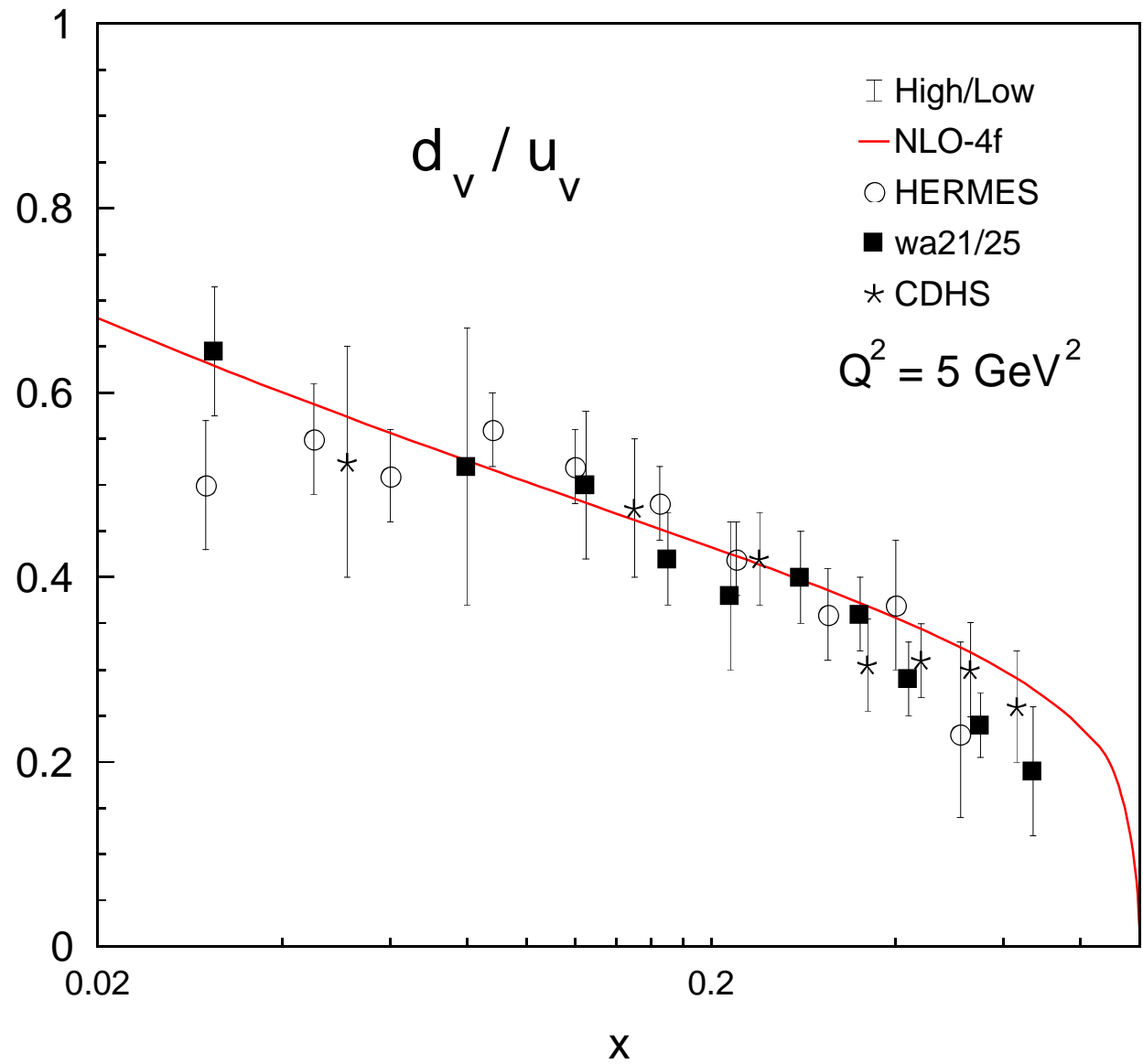
\begin{figure}
\label{fig8}
\caption{
The ratio $\frac{d_{v}}{u_{v}}$ at $Q^{2}=5 GeV^{2}$ evaluated for four flavors.
Data are from CDHS and WA21/25 and Hermes.
}
\end{figure}

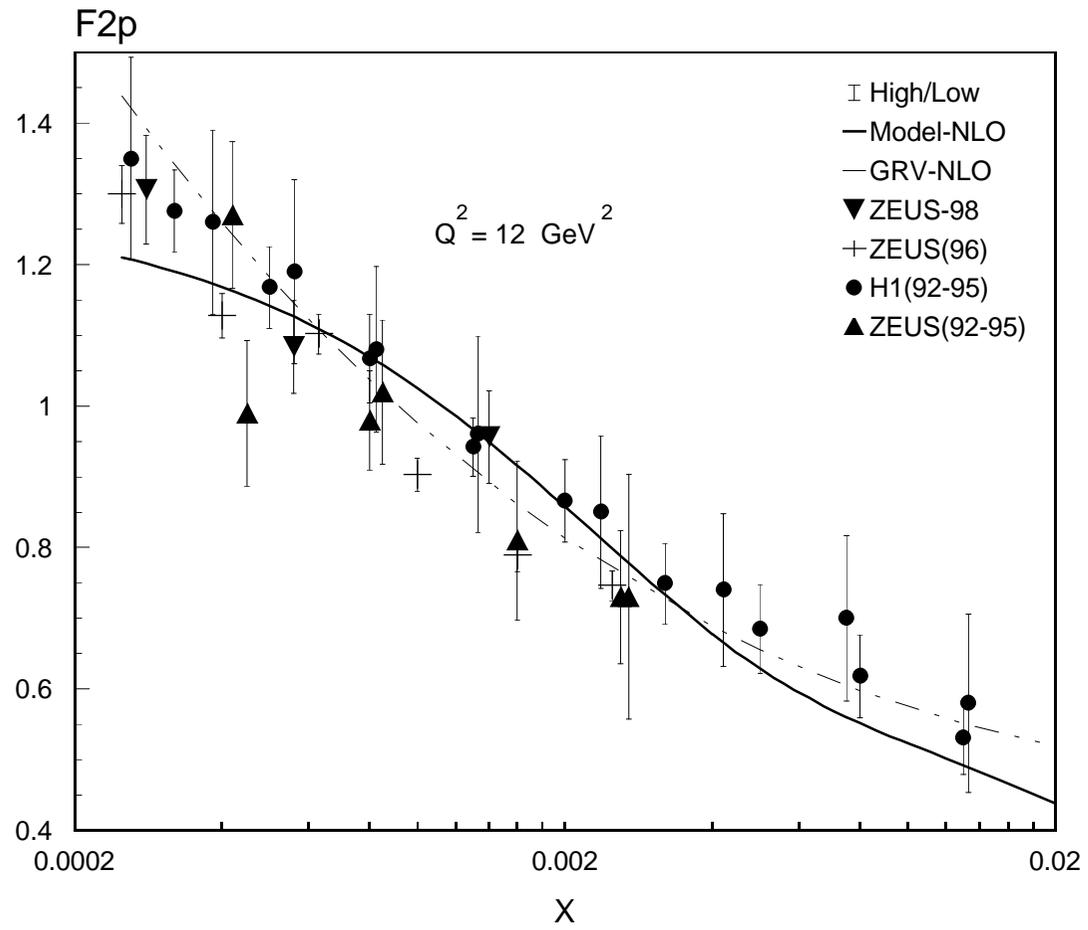
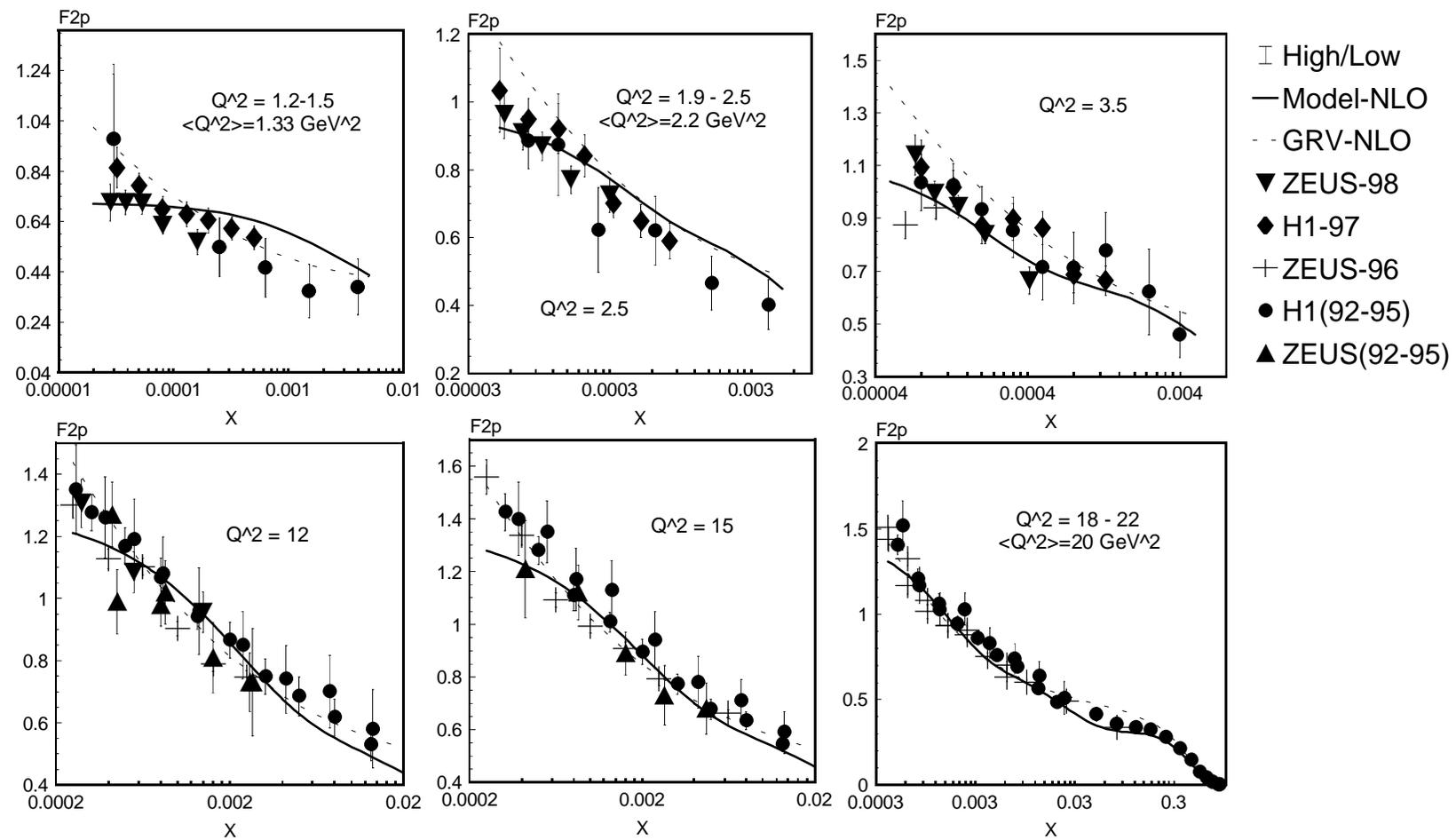
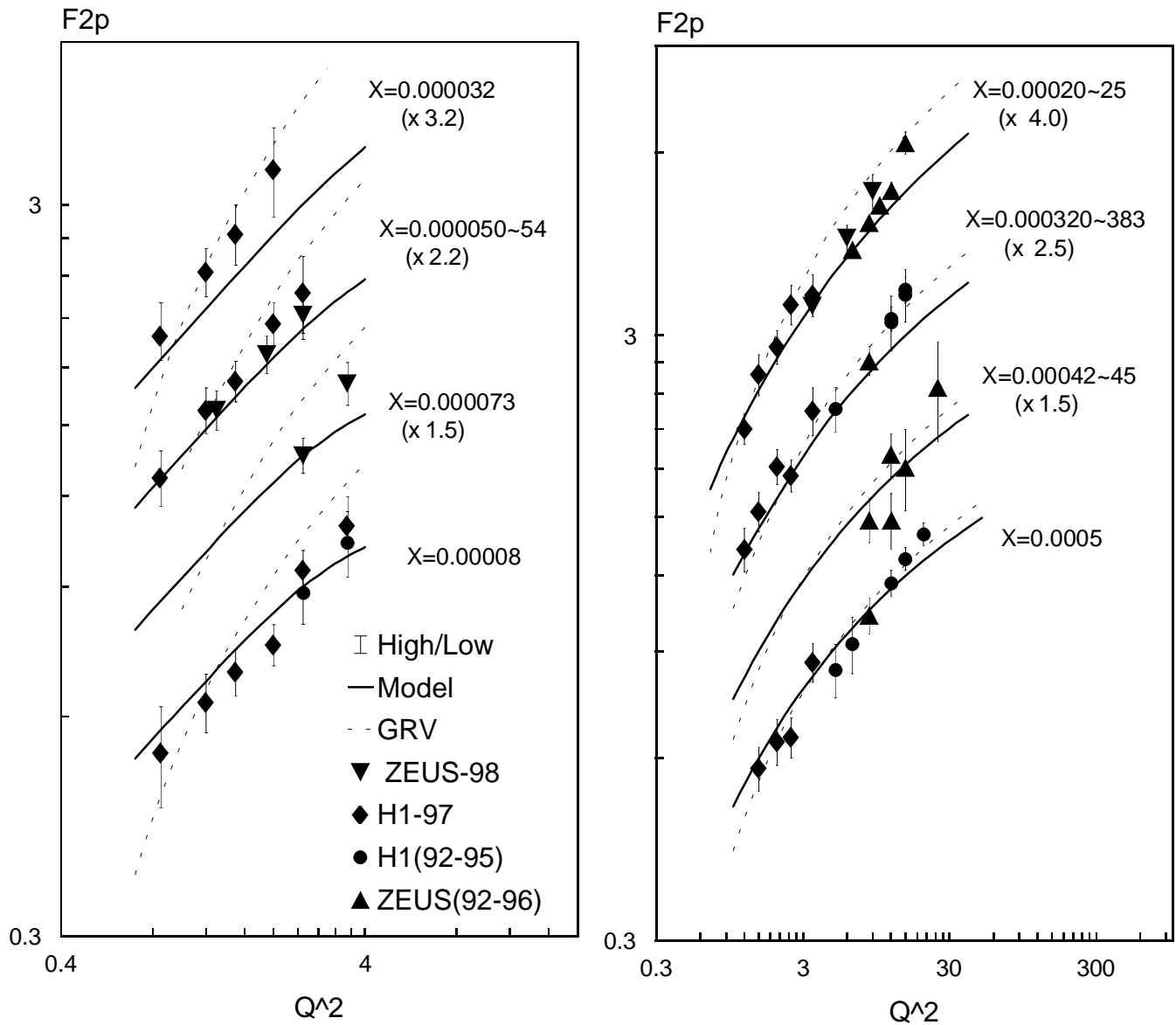
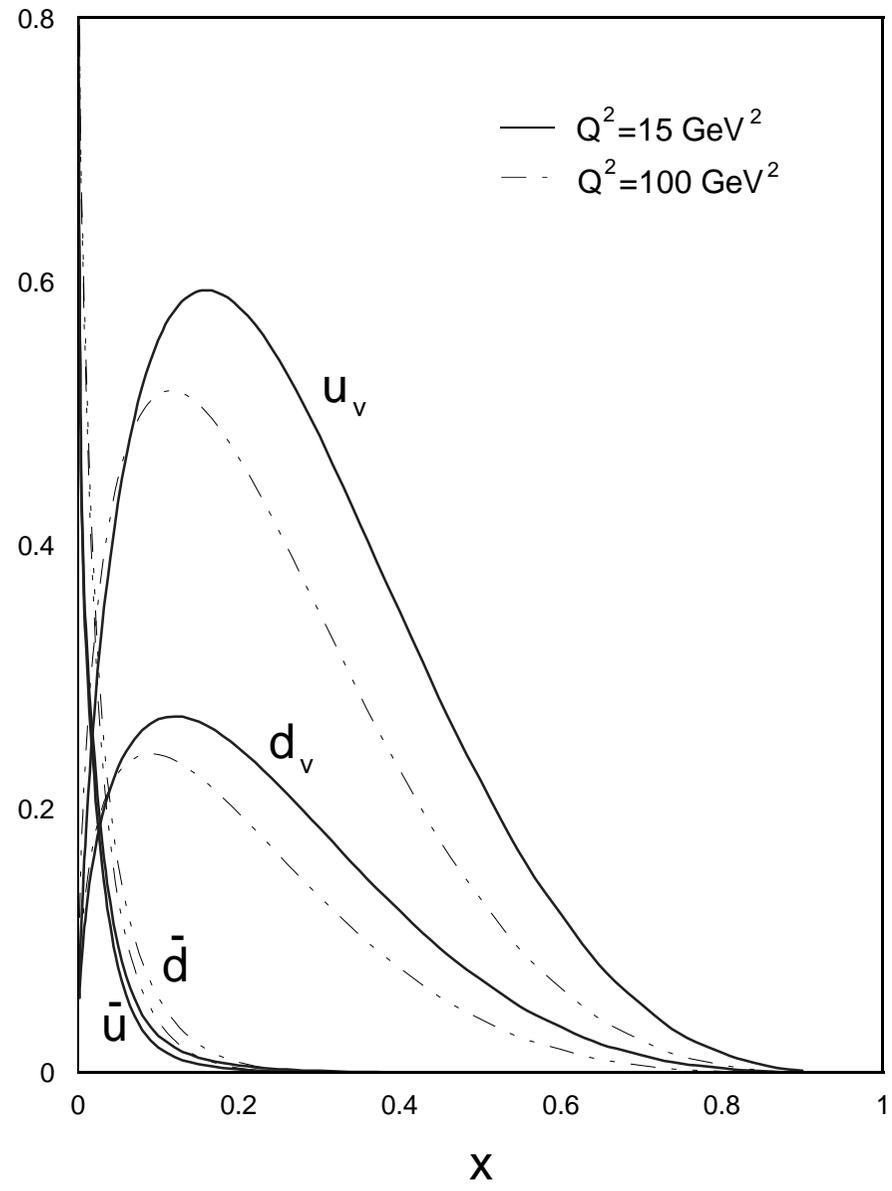
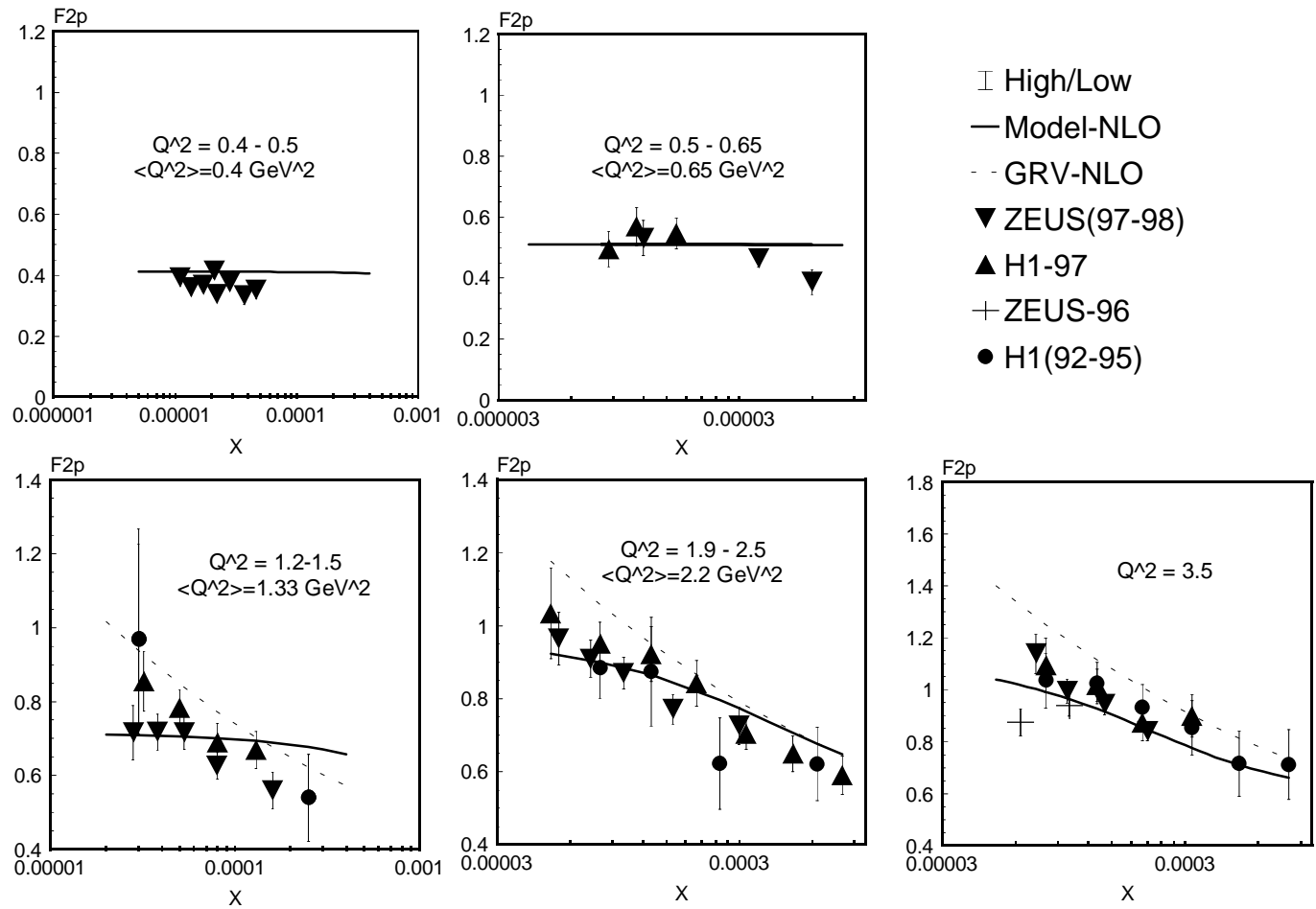
\end{document}